\documentclass[preprint2]{emulateapj} 
\usepackage{graphicx}
\usepackage{graphics}
\usepackage{color}
\usepackage{natbib} 
\DeclareGraphicsExtensions{.pdf}

\begin{document}

\newcommand{\oscdemod}{oscillator/demodulator}
\newcommand{\fMUX}{fMUX}
\newcommand{\squid}{SQUID}
\newcommand{\rtHz}{$\sqrt{\mbox{Hz}}$}
\newcommand{\phinot}{\mbox{$\Phi_0$}}
\newcommand{\degree}{\mbox{$^{\circ}$}}
\newcommand{\fortran}{{\tt Fortran~77}}
\newcommand{\CXX}{C++}
\newcommand{\order}{\mbox{${\cal O}$}}
\newcommand{\const}{\mbox{\sc\small Const}}
\newcommand{\mycomment}[1]{{\bf\it\color{red} #1}} 


\title{Constraints on the High-$\ell$ Power Spectrum of Millimeter-wave Anisotropies from APEX-SZ }

\author{
C.~L.~Reichardt\altaffilmark{1}, 
O.~Zahn\altaffilmark{1},
P.~A.~R.~Ade\altaffilmark{2}, 
K.~Basu\altaffilmark{3},
A.~N.~Bender\altaffilmark{4},
F.~Bertoldi\altaffilmark{5},
H.-M.~Cho\altaffilmark{6}, 
G.~Chon\altaffilmark{3,5,12},
M.~Dobbs\altaffilmark{7}, 
D.~Ferrusca\altaffilmark{1}, 
N.~W.~Halverson\altaffilmark{4,13}, 
W.~L.~Holzapfel\altaffilmark{1},
C.~Horellou\altaffilmark{8},
D.~Johansson\altaffilmark{8},
B.~R.~Johnson\altaffilmark{1},
J.~Kennedy\altaffilmark{7}, 
R.~Kneissl\altaffilmark{3,9,10}, 
T.~Lanting\altaffilmark{2},
A.~T.~Lee\altaffilmark{1,11}, 
M.~Lueker\altaffilmark{1}, 
J.~Mehl\altaffilmark{14}, 
K.~M.~Menten\altaffilmark{3}, 
M.~Nord\altaffilmark{3,5}, 
F.~Pacaud\altaffilmark{5}, 
P.~L.~Richards\altaffilmark{1}, 
R.~Schaaf\altaffilmark{5}, 
D.~Schwan\altaffilmark{1}, 
H.~Spieler\altaffilmark{11}, 
A.~Weiss\altaffilmark{3},
B.~Westbrook\altaffilmark{1}}

\altaffiltext{1}{Department of Physics, University of California, Berkeley, CA, 94720}
\altaffiltext{2}{School of Physics and Astronomy, Cardiff University, CF24 3YB Wales, UK}
\altaffiltext{3}{Max Planck Institute for Radioastronomy, 53121 Bonn, Germany}
\altaffiltext{4} {Center for Astrophysics and Space Astronomy, Department of Astrophysical and Planetary Sciences, University of Colorado, Boulder, CO 80309}
\altaffiltext{5}{Argelander Institute for Astronomy, Bonn University, Bonn, Germany}
\altaffiltext{6}{National Institute of Standards and Technology, Boulder, CO, 80305}
\altaffiltext{7}{Department of Physics, McGill University, Montr\'{e}al, Canada, H3A 2T8}

\altaffiltext{8}{Onsala Space Observatory, Chalmers University of Technology, SE-439 92 Onsala, Sweden}
\altaffiltext{9}{Joint ALMA Office, Av El Golf 40, Piso 18,  Santiago, Chile}
\altaffiltext{10}{ESO, Alonso de Cordova 3107, Vitacura, Santiago, Chile}
\altaffiltext{11}{Lawrence Berkeley National Laboratory, Berkeley, CA, 94720}
\altaffiltext{12}{Max Planck Institute for Extraterrestrial Physics, 85748 Garching, Germany}
\altaffiltext{13}{Department of Physics, University of Colorado, Boulder, CO 80309}
\altaffiltext{14}{University of Chicago, 5640 South Ellis Avenue, Chicago, IL 60637}

\begin{abstract}
We present measurements of the angular power spectrum of millimeter wave 
anisotropies with the APEX-SZ instrument. APEX-SZ has mapped 0.8 square degrees of sky at a frequency of $150\,$GHz with an angular 
resolution of $1^\prime$.
These new measurements significantly improve the constraints on anisotropy power at $150\,$GHz over the range of angular multipoles $3000 < \ell < 10,000$, limiting the total astronomical signal in a flat band power to be less than 105 $\mu$K$^2$ at 95\% CL.
We expect both submillimeter-bright, dusty galaxies and to a lesser extent secondary CMB anisotropies from the Sunyaev-Zel'dovich 
effect (SZE) to significantly contribute to the observed power.
Subtracting the SZE power spectrum expected for $\sigma_8=0.8$ and masking 
bright sources, the best fit value for the remaining power is $C_\ell = 1.1^{+0.9}_{-0.8}  \times10^{-5} \,\mu$K$^2$ ($1.7^{+1.4}_{-1.3}$  Jy$^2$ sr$^{-1}$).
This agrees well with model predictions for power due to submillimeter-bright, dusty galaxies. Comparing this power to the power detected by BLAST at 600 GHz, we find the frequency dependence of the source fluxes to be $S_\nu \propto \nu^{2.6^{+0.4}_{-0.2}}$ if both experiments measure the same population of sources.
Simultaneously fitting for the amplitude of the SZE power spectrum and a Poisson distributed point source population,
we place an upper limit on the matter fluctuation amplitude of $\sigma_8 < 1.18$ at 95\% confidence. 
\end{abstract}


\maketitle

\section{Introduction}

Primary anisotropy in the cosmic microwave background (CMB) radiation is produced 
by inhomogeneities in the hot baryon-photon plasma at the epoch of recombination. 
The amplitude of these anisotropies as a function of angular scale has
been used to infer precise constraints on the parameters of the 
standard $\Lambda{\rm CDM}$ model \citep{dunkley2009,komatsu2008}. 
On angular scales below several arcminutes, the primary anisotropy is
damped by photon diffusion and the observed power is expected to be 
dominated by sources of foreground emission and the interaction
of the CMB with intervening structure.
In particular, the inverse Compton scattering of CMB photons off 
hot plasma bound to clusters of galaxies \citep{sunyaev72} gives rise to
a spectral distortion of the CMB known as the Sunyaev-Zel'dovich effect (SZE). 
At frequencies less than $\sim 220\,$GHz, the SZE  produces a decrement 
in the CMB intensity in the direction of a galaxy cluster. 
Galaxy clusters produce 
(secondary) anisotropy power on arcminute scales corresponding to 
their angular size.
The observed SZE power depends sensitively on the abundance of clusters and 
the history of structure formation.
In particular, the amplitude of the SZE power scales as $\sigma_8^7$, where $\sigma_8$ is the rms fluctuation 
of matter on scales of $8{\rm h}^{-1}\,$Mpc,  and serves as an
independent probe of the amplitude of density perturbations \citep{komatsu2002}.

Evidence for small scale power beyond that expected from the primary CMB 
has been reported by the Berkeley Illinois Maryland
Association (BIMA) and Cosmic Background Imager (CBI) interferometers operating at $30\,$GHz. 
These measurements find a level of SZE anisotropy power consistent with a value of $\sigma_8$ somewhat
greater than those preferred by other contemporary
measurements, which favor $\sigma_8 \sim 0.8$ \citep{vikhlinin2009,komatsu2008}.
BIMA observations at $30\,$GHz covering a total of $0.1\,{\rm deg}^2$ of sky produced a nearly 
$2\sigma$ detection of excess power in a flat band centered at a multipole $\ell=5237$ \citep{dawson2006}.
Due to the non-Gaussian distribution of the SZE on the sky and the low significance of the detection,
this resulted in only weak constraints on the matter power spectrum normalization, 
$\sigma_8=1.03^{+0.20}_{-0.29}$ at $68\%$ confidence.
Observations with the CBI experiment at $30\,$GHz over a larger field were used to produce 
a $> 3\sigma$ detection of excess power on angular scales $\ell \in [1800, 4000]$ \citep{sievers2009}.
Interpreting this power as being due to the SZE, they find $\sigma_8=1.015\pm0.06$ at $68\%$ confidence.  
However, recent observations with the SZA experiment operating at $30\,$GHz and covering a total of 
$2\,{\rm deg}^2$ of sky, 
have determined an upper limit on the excess power in broad band centered at multipole 
$\ell=4066$ which is significantly lower than the band powers reported by CBI and appears to be consistent
with more conventional values of $\sigma_8  \sim 0.8$ \citep{sharp2009}.
The SZA team interprets the higher power measured by CBI as potentially being due to an unsubtracted population 
of radio sources.

At $150\,$GHz, we expect the SZE and emission from distant dusty galaxies to contribute significantly 
to the power on angular scales corresponding to $\ell>2500$.
The ACBAR experiment reported a $\sim 1 \sigma$ excess power at $\ell>2000$, that if interpreted as the SZE, 
would be consistent with the higher value for $\sigma_8$ preferred by CBI \citep{reichardt2008}. 
However, the ACBAR results are in excellent agreement with a more standard value $\sigma_8 \sim 0.8$ 
when one considers the expected foreground emission from IR point sources.
The QUaD collaboration has recently released a $150\,$GHz power spectrum
for $\ell>2000$ \citep{friedman2009}.
The QUaD bandpowers appear (numerical values have not yet been released) to be 
systematically lower than those produced by ACBAR with no evidence for 
contributions from secondary anisotropy or foreground emission. Bolocam has recently published new results for anisotropy power at $150\,$GHz on angular scales above $\ell = 3000$ \citep{sayers09}.  They report upper limits on the power of $1080\,\mu$K$^2$ at 95\% confidence in a wide bin centered at $\ell = 5700$ and determine that $\sigma_8 < 1.57$ at 90\% confidence.
New high resolution and sensitivity bolometer arrays operating at millimeter wavelengths such as 
those currently deployed on the APEX, SPT, and ACT telescopes have the capacity to drastically 
improve constraints on the SZE and point source contributions to the high-$\ell$ power spectrum. 

This paper presents new measurements of small scale anisotropy power made with the 
APEX-SZ bolometer array on the Atacama Pathfinder Experiment (APEX) telescope 
from its high elevation site in the Atacama Desert. This work significantly improves the constraints on excess power above that expected from primary CMB anisotropy at $\ell > 3000$ at a frequency of $150\,$GHz. 
In Section~\ref{SEC:instrumentNObservations}, we describe the APEX-SZ instrument and the
observations used to produce the results presented in this paper.
The beam and calibration of the instrument are described in Section~\ref{SEC:beamcalibration}.
The algorithms used in the production of the temperature maps and the power spectrum
are described in Section~\ref{SEC:analysis}. In Section~\ref{SEC:results}, we present the 
power spectrum results and address sources of foreground emission. Our conclusions
are summarized in Section~\ref{SEC:conclusions}.

\label{SEC:introduction}

\section{Instrument and Observations}
\label{SEC:instrumentNObservations}

APEX-SZ is an array of 330 transition-edge superconducting (TES) bolometers operating at 150 GHz \citep{schwan2003,dobbs2006,schwan2008}.  The bolometers are cooled to 280\,mK via a three stage He
sorption fridge and mechanical pulse tube cooler and instrumented with
a frequency-domain multiplexed readout system.  The array observes from the 12m APEX telescope on the Atacama plateau in Chile \citep{guesten2006} and has approximately 1$^\prime$ FWHM beams with a 22$^\prime$ field-of-view (FOV).  The extremely dry and stable atmospheric conditions make the Atacama one of the best sites for millimeter-wave astronomy.

The band powers reported in this work are derived from a single, 0.8 deg$^2$ field observed by APEX-SZ for 10 nights in August and September of 2007.  This field is a subset of the XMM-LSS field \citep{pierre2004} and is centered on a moderately massive, X-ray detected cluster, XLSSU J022145.2-034614, with an X-ray temperature of 5 keV \citep{willis2005,pacaud2007}. A joint analysis of the X-ray and SZ data will be undertaken in a separate paper. A circular scan strategy was used instead of a raster scan to improve the observing efficiency.  The scan strategy concentrates the integration time at the center of the map causing the time per pixel to increase steadily from the edges of the map to the center.  The total integration time is 2.9k detector-hours. The map center reached a depth of 12 $\mu$K per 1$^\prime$ pixel. More details on the instrument and scan strategy can be found in \cite{halverson08} (hereafter H08) and \cite{schwan2008}.

\section{Beam and Calibration}
\label{SEC:beamcalibration}

The average beam of the APEX-SZ bolometers is measured with daily observations of Mars.  At 8$^{\prime\prime}$ diameter, Mars is nearly a point source for the 1$^\prime$ APEX-SZ beam, and it is sufficiently bright to map the beam near sidelobes to below -25 dB.  The measured beam agrees well with the ZEMAX\footnote{http://www.zemax.com} simulated beam profiles when optical cross-talk is taken into account. The near sidelobes increase the real beam solid angle by 32\% compared to the best fit Gaussian beam. We divide the measurement uncertainty on the beam into two parts.  The main lobe is well-fit by a Gaussian, and we estimate the measurement uncertainty on the FWHM to be 2.5\%. Due to the large angular scale of the sidelobe structure, a mis-estimation of beam sidelobe will effectively cause a mis-calibration of the band powers. We include the uncertainty in the total beam area in the calibration error.

The observations of Mars are used to establish the absolute calibration
of the APEX-SZ instrument. 
The temperature of Mars for our observation frequency and dates is taken 
from the Rudy model \citep{rudy87,muhleman91}, that has been updated and 
maintained by 
Bryan Butler\footnote{http://www.aoc.nrao.edu/$\sim$bbutler/work/mars/model/}. 
The Rudy model is compared with measurements of the brightness 
temperature of Mars made with the WMAP satellite at $93\,$GHz during 
five periods across several years \citep{hill08}.
The WMAP measurements of Mars have uncertainty of $<1\%$ and we adjust the
normalization of the Rudy Model down by $5.2\%$ to bring it into agreement 
with the WMAP measurements.
Combining the $\sim 1\%$ uncertainty in the WMAP Mars measurements, the $1.0\%$
scatter in the $93\,$GHz WMAP to Rudy Model comparison, 
and $0.9\%$ for the uncertainty in the extrapolation from $93\,$GHz, 
where the Rudy Model is calibrated, to our observing frequency, we find 
the total uncertainty in the Mars brightness temperature to be $1.7\%$.

The calibration of each detector is set by comparing the peak amplitude in a map of Mars to the expected amplitude given the temperature and size of Mars and the size of the detector's beam. 
A correction factor for the atmospheric opacity is applied which is always 
less than 3\%. 
The overall calibration uncertainty is estimated to be 5.5\% in temperature, 
with the dominant errors due to a 4\% uncertainty in the beam area, 
a 3\% uncertainty to account for temporal variations between the model prediction and the temperature measured by APEX-SZ, the 1.7\% uncertainty in the temperature 
of Mars, and a $1.4\%$ uncertainty in the conversion of brightness 
temperature to CMB temperature for the measured APEX-SZ band.  
More details on the beam estimation and calibration can be found in H08.

\section{Analysis}
\label{SEC:analysis}

\subsection{Map-Making}
\label{SUBSEC:mapmaking}

The filtering and map-making process used in this analysis follows the approach detailed in H08 for analysis of the Bullet cluster.  We briefly outline the major steps here, while highlighting any differences in the filtering between this work and the Bullet cluster analysis.  The timestream processing is designed to remove scan-synchronous noise, atmospheric fluctuations, and 1/f noise.  The scan pattern includes boresight elevation changes which modulate the air mass along the line of sight.  This signal is removed by fitting for a cosecant($el$) term.  1/f noise in the system is filtered from the timestream by a 0.3 Hz 8-pole Butterworth high-pass filter (HPF).  The typical length scale of atmospheric fluctuations is much larger than the FOV, so fluctuations are highly correlated across the APEX-SZ array.   This correlated term is removed by fitting for a low order spatial polynomial across the focal plane at each time sample.  We remove a 2$^{nd}$ order spatial mode in this work. The cumulative effect of the filtering is to completely remove structures corresponding to angular multipoles below $\ell \simeq 1400$. After filtering, the timestreams are weighted according to the filtered timestream RMS and binned into 20$^{\prime\prime}$ map pixels.

\subsection{Power Spectrum Estimation}
\label{SUBSEC:psestimate}

The band powers, $q_B$, are reported in CMB temperature units of $\mu$K$^2$ and parametrize the power spectrum according to
   \begin{equation}  \frac{\ell (\ell + 1)}{ 2 \pi} C_\ell  \equiv {\it D}_\ell = \sum_B q_B \chi_{B\ell} 
   \end{equation}
where $\chi_{B\ell}$ are top hat functions; $\chi_{B\ell} = 1$ for $\ell \in B$ and 0 for $\ell \not\in B$. We use a pseudo-$\it{C}_\ell$ power spectrum estimator \citep{hivon02}.  In this formalism, the map spectrum (also called the pseudo-$C_\ell$ or $\tilde{C}_\ell$) depends on the true spectrum ($C_\ell$) as:
     \begin{equation} \tilde{C}_\ell = M_{\ell \ell^\prime} T_{\ell^\prime} B^2_{\ell^\prime} C_{\ell^\prime}.
     \end{equation}
  $\tilde{C}_\ell$ is calculated using the flat-sky approximation, in which the spherical harmonic transform of the sky reduces to a Fourier transform of the map. This is an excellent approximation for a sub-degree sized map.  The experimental beam function is described by $B_\ell$, and the mode-coupling due to finite sky coverage is denoted by $M_{\ell \ell^\prime} $. $T_\ell$ represents the transfer function of the map-making process which would ideally be equal to one. In practice, the HPF applied to the APEX-SZ timestreams eliminates power on scales $\ell < 1400$ so $T_\ell$ = 0 on these scales and remains below one at all $\ell$.  Following the MASTER algorithm \citep{hivon02}, we measure the transfer function using a set of Monte-Carlo sky realizations  that have been passed through the full pipeline, from the timestreams to the maps. These simulated sky maps include a lensed WMAP5+ACBAR best fit CMB model \citep{hill08,reichardt2008}, realizations of the SZE signal \citep{shaw2009}, and realizations of the point source populations \citep{negrello2007,granato2004,zotti2005}. This set of signal-only simulations is also used to estimate the cosmic variance contribution to the band power uncertainties. 
  
  The mode-coupling matrix $M_{\ell \ell^\prime}$ is calculated analytically for the two sky masks used to estimate the APEX-SZ band powers. These masks describe the weighting applied to each pixel in the map before calculating the Fourier transform and are analogous to windowing data in a 1D Fourier transform. We begin by applying an inverse-noise weighting to each pixel, based on the diagonal elements of the pixel-pixel noise covariance matrix. This effectively de-weights the noisy edges of the map, and is near-optimal in the low signal-to-noise per pixel regime.  We modify this simple mask to exclude pixels near detected clusters or point sources. There are two X-ray detected clusters in the field; XLSSU J022145.2-034614 and XLSSU J022157.4-034001. The field was centered on the first and more massive of these clusters, which would introduce a bias into the determination of the SZE amplitude. The central cluster is masked to a diameter of 6$^\prime$ in the first mask, hereafter the {\it Cluster} mask.  The second cluster is fainter, and was detected only in a joint analysis of X-ray and optical data \citep{andreon2005}. We tested the effects of masking the second cluster as well and did not see a significant change in the band powers. The second mask removes the 27 point sources detected in the APEX-SZ map in addition to the central cluster. These sources are found by using SExtractor \citep{bertin1996} to select sets of neighboring pixels above 3$\,\sigma$ in an optimally filtered map. This corresponds approximately to a detection threshold of 2 mJy. We discuss these sources further in \S\ref{SEC:source_contributions}.  This mask will be referred to as the {\it Cluster + Sources} mask. We tested the effect of masking all NVSS or bright Spitzer sources within the field, and observed no significant change in power.
  
   The measured band powers will be the sum of the signal and noise band powers,
   \begin{equation} 
    {\it D}_\ell  = {\it D}_\ell^S  + {\it D}_\ell^N 
   \end{equation}
  and we must subtract the expected noise contribution to recover the underlying signal spectrum. The noise contribution to the APEX-SZ band powers is estimated from the average power in a set of 2200 jack-knife maps between two randomly selected half sets of the $\sim$1100 complete observations of the field. These difference maps effectively remove sky signal, while preserving correlated noise in the timestreams on time scales shorter than the few minute length of a scan with randomized phase. Noise on longer time scales has been removed by the 0.3 Hz HPF. The expectation value of the noise band powers, $\langle{\it D}_\ell^N \rangle$, is taken to be the mean band powers measured across the set of 2200 jack-knives. The approach is similar to that used in the analysis of the Bolocam power spectrum \citep{sayers09}. We apply the same procedure to signal-only simulated maps and confirm that any residual signal power due to the small pointing, filtering and weighting differences between observations is negligible.

\section{Band powers and $\sigma_8$}
\label{SEC:results}
\begin{figure*}[ht]\centering
\includegraphics[width=0.9\textwidth]{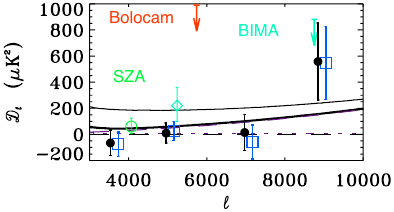}
  \caption[]{
  Band powers derived from the APEX-SZ map plotted over a model ({\bf thick black line}) including the primary CMB anisotropies, a SZE model for $\sigma_8$ = 0.8 ({\bf short dashed purple line}) , and the predicted point source contribution for a 2 mJy cut threshold ({\bf long dashed purple line}) . This model is not a fit to the APEX-SZ band powers. We also plot for comparison the theory spectrum ({\bf thin black line}) if we increase $\sigma_8$ to APEX-SZ's 95\% CL upper limit of 1.18.
  The APEX-SZ band powers for the {\it Cluster} mask are plotted with {\bf blue squares}, while the {\it Cluster+Sources} mask results are shown as {\bf black circles}. The  {\it Cluster} mask band powers have been shifted by $\Delta\ell = 200$ to the right for clarity.  BIMA ({\bf turquoise diamond and upper limit} - \cite{dawson2006}), SZA ({\bf green circle} - \cite{sharp2009}), and Bolocam ({\bf red upper limit} - \cite{sayers09}) have previously released band-powers centered at $\ell >$ 3000.  The upper limits are shown at 95\% CL. BIMA and SZA operate at 30 GHz where there will be four times as much SZE power as at 150 GHz and we expect the foregrounds to be dominated by radio sources rather than dusty galaxies. The plotted theory spectra are for 150 GHz only.}
\label{fig:spectrum}
\end{figure*}

\begin{table*}[ht!]
\begin{center}
\caption{Band powers} 
\small
\begin{tabular}{cc||cc|cc}
\hline\hline
\rule[-2mm]{0mm}{6mm}
 & &\multicolumn{2}{c}{Cluster masked}&\multicolumn{2}{c}{Cluster + Sources masked} \\
$\ell$ range&$\ell_{\rm eff}$ &$q$ ($\mu{\rm K}^2$)& $\sigma$ ($\mu{\rm K}^2$) &$q$ ($\mu{\rm K}^2$)& $\sigma$ ($\mu{\rm K}^2$)\\
\hline

3000-4000 & 3532      &  -74& 94    & -66&98\\
4000-6000 & 4957       & 26 &73     &9 &77\\
6000-8000 & 6968  & -58& 132  & 14&138\\
8000-10000 & 8844 & 546& 280  & 558&296\\

\hline
\end{tabular}

  \tablecomments{\small
Band multipole range and weighted value $\ell_{\rm eff}$, band powers $q$, and
uncertainty $\sigma$ from the analysis of the XMM-LSS field with two masks. The first two columns ({\it Cluster masked}) show the results when the central X-ray detected cluster in the 
field is masked to 6$^\prime$ diameter. 
The second set of columns ({\it Cluster + Sources masked}) masks twenty-seven $>3 \sigma$ sources detected in the map to 1.5$^\prime$ diameter as well as the cluster. More details on the masks can be found in \S\ref{SUBSEC:psestimate}.
}
\label{tab:bands}
\normalsize
\end{center}
\end{table*}

The power spectrum presented in Figure \ref{fig:spectrum} is the product of applying the analysis in Section \ref{SEC:analysis} to APEX-SZ observations of the XMM-LSS field. The band powers for angular multipoles from 3000 to 10,000 are tabulated in Table \ref{tab:bands}. The band powers can be compared to a theoretical model using the window functions. The numerical values for the band powers and window functions can be downloaded from the APEX-SZ website\footnote{http://bolo.berkeley.edu/apexsz/index.html}.

\begin{table*}[ht!]
\begin{center}
\caption{Point source power and $\sigma_8$ constraints} 
\small
\begin{tabular}{c|cc}
\hline\hline
\rule[-2mm]{0mm}{6mm}
 &  Cluster masked& Cluster + Sources masked\\
 \hline\hline
Zero SZE power: &  &  \\
$\it{C}_\ell^{PS}$ ($10^{-5}$ $\mu{\rm K}^2)$& $1.0^{+0.9}_{-0.6} $& $1.2^{+1.0}_{-0.8}  $\\
\hline
 Fixed $\sigma_8=0.8: $&  &  \\
$\it{C}_\ell^{PS}$ ($10^{-5}$ $\mu{\rm K}^2)$  & $0.9^{+0.9}_{-0.6} $&$1.1^{+0.9}_{-0.8}  $\\
\hline
 Unconstrained $\sigma_8$:&  &  \\
$\it{C}_\ell^{PS}$ ($10^{-5}$ $\mu{\rm K}^2)$ & $0.9^{+0.9}_{-0.6} $&$1.1^{+0.9}_{-0.8} $\\
$\sigma_8$ (G) (95\% CL) &  0.94& 0.94 \\
$\sigma_8$ (NG) (95\% CL) &  1.18& 1.18 \\
\hline
\hline
 Flat excess: & & \\
 (with $\ell_{center}$ = 4966)& & \\
  $\it{D}_\ell$ ($\mu{\rm K}^2$) & $33^{+37}_{-24}$ & $36^{+39}_{-26}$\\
  95\% CL & 97 & 105 \\
\hline
\end{tabular}
  \tablecomments{\small	
The constraint on point source power $\it{C}_\ell^{PS}$, and the 95\% CL upper limit on $\sigma_8$ derived from the APEX-SZ data set are tabulated for different assumptions about the SZE power and masks. The expected primary CMB anisotropy power has been subtracted from the measured band powers.  We show the upper limit on $\sigma_8$ with (NG) and without (G) accounting for the non-Gaussian distribution of the SZE. Accounting for the non-Gaussianity in the expected SZE sky weakens the upper limit considerably.
The first column ({\it Cluster masked}) shows the results when the massive, X-ray detected cluster in the 
field is masked to 6$^\prime$ diameter. The second column 
({\it Cluster + Sources masked}) excludes the $>3 \sigma$ sources detected in the map as well as the cluster. 
More details on the masks can be found in \S\ref{SUBSEC:psestimate}. 
The measured point source power is in excellent agreement with the predicted amplitude 
of $1.1 \times 10^{-5}\, \mu$K$^2$ (1.7 Jy$^2$ sr$^{-1}$) for the sum of the radio and dusty galaxy models 
in all cases. We also show the results under the assumption that the power at high-$\ell$ above the primary CMB anisotropies can be modeled as a flat band-power. The 95\% CL upper limit on a flat excess of $<$105 $\mu{\rm K}^2$ includes the beam and calibration uncertainties but does not include non-Gaussian contributions to cosmic variance.
}
\label{tab:params}
\normalsize
\end{center}
\end{table*}

\begin{figure*}[ht]\centering
\includegraphics[width=0.9\textwidth]{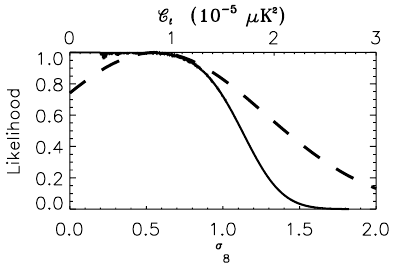}
  \caption[]{ [{\it \b solid and lower x-axis}] Marginalized likelihood function for $\sigma_8$ for a flat prior on $\sigma_8$ after accounting for non-Gaussianity. [{\it \b dashed and upper x-axis}] Marginalized likelihood function for $C^{ps}_\ell$. These curves are for the {\it Cluster+Sources} mask. }
\label{FIG:like1d}
\end{figure*}

\begin{figure*}[t]\centering
\includegraphics[width=0.8\textwidth]{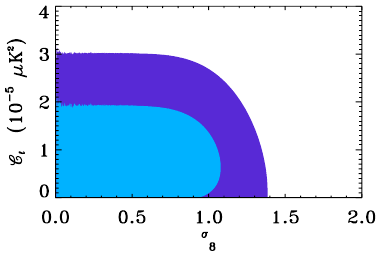}
  \caption[]{ The 1-$\sigma$ and 2-$\sigma$ contours on the 2D likelihood surface for $\sigma_8$ and the point source amplitude for the {\it Cluster+Sources} mask applied to the APEX-SZ map. The contours include the effects of non-Gaussianity in the SZE. A flat prior on $\sigma_8$ has been assumed. }
\label{FIG:like2d}
\end{figure*}

The APEX-SZ band powers show a tendency to increase on smaller angular scales, suggestive of 
the $\ell^2$ shape that a Poisson distribution of point sources will have in a plot of $\ell (\ell + 1) C_\ell/ 2 \pi$.  
The contribution of the primary CMB will be small at these small angular scales, and we subtract the 
estimated contribution for the best fit WMAP5+ACBAR lensed $\Lambda{\rm CDM}$ power spectrum before fitting 
for the amplitude of a constant $C_\ell$. The beam and calibration uncertainty is incorporated by 
averaging the likelihood function over a set of 200 Monte Carlo realizations. 
The best fit power is $C_\ell = 1.0^{+0.9}_{-0.6}  \times10^{-5} \,\mu{\rm K}^2$, which includes both the point 
source and SZE contributions.  Results for both masks are tabulated in the first row of Table \ref{tab:params}.
If the expected SZE power spectrum for $\sigma_8 = 0.8$ is subtracted in addition to the primary 
anisotropies, the average $C_\ell$ drops slightly to $0.9^{+0.9}_{-0.6} \times10^{-5} \,\mu {\rm K}^2$. At the best-fit point source amplitude for $\sigma_8 = 0.8$, 92\% of the astronomical power in the map is produced by point sources rather than the SZE.  These results are obtained after masking the bright, central cluster. Leaving the central cluster unmasked increases the fit power by $\sim$1.0 $ \times10^{-5}\, \mu {\rm K}^2$.
We also report the results after masking the sources internally detected 
in the map at $> 3 \sigma$ in the second row of Table \ref{tab:params}. These results are consistent with and improved over the previous ACBAR constraints from $\ell < 3000$ of $C_\ell = 2.7^{+1.1}_{-2.6}  \times10^{-5} \,\mu {\rm K}^2$. However, these numbers should be compared with caution as the two experiments have different flux cuts for source masking, and the excess power will depend on the flux to which sources have been masked.

We also investigate the effects of allowing the amplitude of the SZE power spectrum to float freely.  The SZE power spectrum template is based on the simulations in \cite{shaw2009}. The simulations are for a WMAP5 cosmology with $\sigma_8 = 0.77$. 
The amplitude of the SZE power spectrum is expected to scale approximately as $\sigma_8^7$, so the derived SZE amplitude can be related to $\sigma_8$. In practice, the APEX-SZ data set lacks the sensitivity to make a detection of SZE power; however, the results can be used to place an upper limit on $\sigma_8$.
The upper limits on $\sigma_8$ and the point source amplitudes for the joint fit are reported in Table \ref{tab:params}. We assume a flat prior on $\sigma_8$.
The exact amplitude of the SZE spectrum is only poorly understood, leading to a 10\% systematic uncertainty on $\sigma_8$ \citep{komatsu2002}.
This systematic uncertainty is not included in the reported upper limits.

The non-Gaussian distribution of the SZE is very important on small patches of sky \citep{cooray2001,zhang2006}. We incorporate the non-Gaussian statistics into our analysis by returning to the set of simulated SZ skys \citep{shaw2009}.  We extract 7500 independent realizations of the APEX-SZ map and calculate the power in each realization under the two masks. The maps have been convolved by the experimental beam and do not include noise. This process maps out the full, non-Gaussian cosmic variance of the expected SZE power for $\sigma_8 = 0.77$.   We scale this to other cosmologies by assuming that the probability of measuring a power X will scale with $\sigma_8$ as 
\begin{equation}P(X | \sigma_8) = \left( \frac{\sigma_8}{0.77} \right)^7 P( \left( \frac{0.77}{\sigma_8} \right)^7 X\,| \sigma_8=0.77 ).\end{equation}
 Bayes' theorem with a flat prior in $\sigma_8$ is applied to find a posterior probability density,  $P(\sigma_8| X)$. 
We determine the probability of a given SZE power from the data by marginalizing over a point source component as described in the last paragraph. Finally, the likelihood function of $\sigma_8$ given the APEX-SZ data ${\bf d}$ is calculated by
   \begin{equation} P(\sigma_8 | {\bf d}) = \int{dX P(\sigma_8 | X) P(X | {\bf d} ) }
   \end{equation}
and integrated to find the 95\% CL upper limit on $\sigma_8$.  The limit rises to $\sigma_8 < 1.18$, substantially weaker than the limit of $\sigma_8 < 0.94$ derived under Gaussian assumptions (see the third row of Table \ref{tab:params}, labeled ``Unconstrained $\sigma_8$"). The marginalized likelihood function for both $\sigma_8$ and $C_\ell^{PS}$ are plotted in Figure \ref{FIG:like1d}, while the 2d likelihood surface for both parameters is shown in Figure \ref{FIG:like2d}. The upper limit is sensitive to the prior chosen since APEX-SZ does not make a detection of SZE power.  A flat prior on power, $\sigma_8^7$, strongly prefers higher values of $\sigma_8$ than the flat prior on $\sigma_8$, and raises the upper limit from 1.18 to $\sigma_8 < 1.50$ at 95\% CL. This is entirely due to the weighting by the prior as the prior probability for $\sigma_8 = 2$ is 240 times the probability of $\sigma_8=0.8$. 

Finally, we combine the four APEX-SZ band powers into a single band to facilitate the comparison to other data sets. The resulting upper limit is $\sim 100\, \mu$K$^2$ after including the APEX-SZ calibration and beam uncertainty as shown in the last row of Table \ref{tab:params}, ``Flat Excess". We have assumed that ${\it D}_\ell$ is constant across the four bands and subtracted the contribution due to the primary CMB anisotropies. However, this upper limit does not include a non-Gaussian contribution to cosmic variance.

\subsection{Radio and IR source contributions}
\label{SEC:source_contributions}

Submillimeter bright galaxies and radio sources are expected to dominate the primary temperature anisotropies for $\ell \gtrsim 2500$ at 150 GHz.  The exact contribution from point sources, especially radio sources, will depend on our ability to detect and mask the brightest sources. The exact 3$\,\sigma$ detection threshold in the APEX-SZ map depends on the map position due to the uneven coverage, but is approximately 2 mJy on average. As discussed below, the predicted band powers are fairly insensitive to the precise cut level unless it shifts by an order of magnitude. We assume both populations are drawn from a Poisson distribution.

The number counts of dusty, submillimeter bright galaxies are modeled in \citet{negrello2007} based on surveys at higher frequencies.  Deep, high-resolution maps of the 150 GHz sky are expected to be confusion-limited by these sources. The anisotropies are the result of variations in the number of very faint sources with fluxes around 0.5 mJy. A 1 mJy point source will produce an increment of  20 $\mu$K in the APEX-SZ map. The APEX-SZ map of the XMM-LSS field is too shallow to pick out these sub-mJy sources, and we see no evidence of reaching the confusion-limit in the current observations.
The dusty galaxy contribution to the APEX-SZ band powers is predicted by the model presented in \citet{negrello2007} to be $C_\ell$ = 1.1 $\times 10^{-5}$ $\mu$K$^2$ (1.7 Jy$^2$ sr$^{-1}$) in the absence of clustering, which is in good 
agreement with the measured point source power in Table \ref{tab:params}.  The predicted power from dusty galaxies is nearly independent of the flux cut level above 1 mJy.
Dusty galaxies are expected to account for most of the power in the APEX-SZ maps.  

The BLAST collaboration recently released measurements of the power spectrum of the cosmic far-infrared background at frequencies of 600 GHz to 1.2 THz \citep{viero2009}.  BLAST measured a Poisson contribution from star-forming galaxies with an amplitude of $2.63 \pm 0.1 \times 10^3$ Jy$^2$ sr$^{-1}$ at 600 GHz. A clustering term is detected as well on angular scales larger than those probed by APEX-SZ.  Expressing the frequency dependence of the source fluxes as $S(\nu) \propto \nu^\alpha$, we can derive an effective spectral index, $\alpha$, by comparing the power measured by BLAST at 600 GHz to the point source power likelihood function of the APEX-SZ maps at 150 GHz.  This index will depend on the the spectra of the individual galaxies and their redshift distribution. We find that a spectral index of $\alpha=2.64^{+0.4}_{-0.2}$ scales the BLAST power to match the best-fit $C_\ell$ = 1.1$^{+0.9}_{-0.8} \times 10^{-5}$ $\mu$K$^2$ (1.7$^{+1.4}_{-1.3}$ Jy$^2$ sr$^{-1}$) of the APEX-SZ data.  This inferred spectral index agrees well with previous estimates for sub-mm bright galaxies. \citet{knox2004} examined nearby galaxy data and found $S_\nu \propto \nu^{2.6 \pm 0.3}$.  In an alternative approach, \citet{greve2004} compared the flux of sources in overlapping regions observed by MAMBO (1.2 mm) and SCUBA (850 $\mu$m) and found the fluxes scaled as $S_\nu \propto \nu^{2.65}$. The point source power in the APEX-SZ data set at 150 GHz is consistent with being entirely due to a population of dusty submm-bright galaxies such as those observed by BLAST.

We also consider radio sources as a potential foreground in the APEX-SZ maps. \cite{granato2004} and \cite{zotti2005} have modeled the number counts of several classes of radio sources at tens of GHz. 
We derive ${\it C}_\ell^{\rm radio}$ from their modeled number counts at 150 GHz.  
The radio source power is dependent on the brightest objects and is expected to scale approximately 
linearly with the source cut threshold. 
At the 2 mJy source cut threshold of APEX-SZ, ${\it C}_\ell^{\rm radio}$ should be $\lesssim 5\%$ of the dusty galaxy contribution.

We find 27 point sources above 3 $\sigma$ ($\sim$2 mJy) in the APEX-SZ maps using the approach outlined in \S\ref{SUBSEC:psestimate}.  Eight of these sources are within 1$^\prime$ of a NVSS source and are likely radio sources.  We expect four false detections based on Gaussian statistics and the number of beam-sized pixels in the APEX-SZ map. The remaining sources are tentatively identified as dusty galaxies. We can compare the observed number counts in the APEX-SZ maps with other experiments at 150 GHz, however most previous experiments were targeting larger angular scales and are relatively insensitive to dim point sources.  QUaD \citep{friedman2009} and ACBAR \citep{reichardt2008} both report $\sim$0.1 radio sources per square degree with a flux detection threshold of many tens of mJy. The deepest previously published map at 150 GHz is from Bolocam \citep{sayers09} which reports no sources above 10 mJy in a 1 deg$^2$ patch. As dN($>$S)/dS is expected to fall steeply above 1 mJy for dusty galaxies, the scarcity of detected sources in these maps is not particularly surprising.  The BLAST source catalogue \citep{dye2009} of 351 sources detected at 250, 350 or 500 $\mu$m allows a more interesting cross-comparison. The BLAST catalogue has 294 sources per square degree in the deep coverage region and 15 sources per square degree in the shallow coverage region. At 500 $\mu$m, the detection threshold of the BLAST catalogue is 30 mJy and 100 mJy respectively. For the best-fit effective spectral index of $\alpha=2.64$ derived earlier, this corresponds to a source detection threshold at 150 GHz of 0.8 and 2.6 mJy respectively. These two detection thresholds bracket the average APEX-SZ 3$\,\sigma$ detection threshold of 2 mJy, and the observed APEX-SZ non-radio source number density of 19 sources/deg$^2$ falls between these two source densities as we would expect. The source number counts in the APEX-SZ maps appear consistent with the numbers expected for dusty galaxies.


\section{Conclusions}
\label{SEC:conclusions}

Observations with the APEX-SZ instrument have been used to constrain the power in excess of the primary 
CMB temperature anisotropies at $150\,$GHz.  
This power is expected to be dominated by emission from sub-mm bright, dusty galaxies and the APEX-SZ 
band powers are consistent with this hypothesis. 
We find excellent agreement between the point source power in the APEX-SZ maps and model predictions based on observations at other frequencies. We estimate that the flux of these sub-mm bright galaxies scales with frequency as $S_\nu \sim \nu^{2.64}$ by comparing the power measured by BLAST at 600 GHz to the best-fit point source power in the APEX-SZ maps at 150 GHz.
Determining the contribution of these foreground sources not only constrains models for the population of 
dusty galaxies, but is important for planning current and future observations of secondary CMB anisotropies
at these wavelengths.

We also place upper limits on $\sigma_8$ from fits to the amplitude of the SZE power spectrum while 
marginalizing over a Poisson point source contribution. 
We assume a template for the SZE power spectrum derived from simulations by \cite{shaw2009} with  
the amplitude of the SZE power spectrum scaling as $\sigma_8^7$. 
We find an upper limit of $\sigma_8< 1.18$ at 95\% confidence.  
This result is similar to the constraints from the CBI and BIMA interferometers operating 
at $30\,$GHz.
The limits from SZA, QUaD, or ACBAR would likely be slightly lower, but they did not express their results 
in terms of upper limits on $\sigma_8$.
At these frequencies and angular scales, the previous best limit comes from \citet{sayers09}, who used
observations with the Bolocam instrument to constrain $\sigma_8<1.57$ at $90\%$ confidence.

A third of the APEX-SZ instrument was recently upgraded to more sensitive detectors with improved optical efficiencies. 
In the next year, the remainder of the focal plane will be upgraded resulting in significant improvements to the instrument's mapping speed.
The instrument will continue observing in the next several years with a focus on developing a 
catalog of clusters with SZE, X-ray and optical observations across the Southern hemisphere.

\section{Acknowledgments}

We thank the staff at the APEX telescope site, led by David
Rabanus and previously by Lars-\AA ke Nyman, for their dedicated and exceptional support. We also thank
LBNL engineers John Joseph and Chinh Vu for their work on the
readout electronics. APEX-SZ is funded by the National Science
Foundation under Grant No.\ AST-0138348. Work at LBNL is supported
by the Director, Office of Science, Office of High Energy and
Nuclear Physics, of the U.S. Department of Energy under Contract No.
DE-AC02-05CH11231. Work at McGill is supported by the Natural Sciences
and Engineering Research Council of Canada and the Canadian Institute
for Advanced Research. This research used resources of the National Energy Research Scientific Computing Center, which is supported by the Office of Science of the U.S. Department of Energy under Contract No. DE-AC02-05CH11231.  Nils Halverson acknowledges support from an Alfred P. Sloan Research Fellowship.

\bibliographystyle{apj}
\bibliography{master_references}

\begin{thebibliography}{35}
\expandafter\ifx\csname natexlab\endcsname\relax\def\natexlab#1{#1}\fi

\bibitem[{{Andreon} {et~al.}(2005){Andreon}, {Valtchanov}, {Jones}, {Altieri},
  {Bremer}, {Willis}, {Pierre}, \& {Quintana}}]{andreon2005}
{Andreon}, S., {Valtchanov}, I., {Jones}, L.~R., {Altieri}, B., {Bremer}, M.,
  {Willis}, J., {Pierre}, M., \& {Quintana}, H. 2005, \mnras, 359, 1250

\bibitem[{{Bertin} \& {Arnouts}(1996)}]{bertin1996}
{Bertin}, E. \& {Arnouts}, S. 1996, \aaps, 117, 393

\bibitem[{{Cooray}(2001)}]{cooray2001}
{Cooray}, A. 2001, \prd, 64, 063514

\bibitem[{{Dawson} {et~al.}(2006){Dawson}, {Holzapfel}, {Carlstrom}, {Joy}, \&
  {LaRoque}}]{dawson2006}
{Dawson}, K.~S., {Holzapfel}, W.~L., {Carlstrom}, J.~E., {Joy}, M., \&
  {LaRoque}, S.~J. 2006, \apj, 647, 13

\bibitem[{{de Zotti} {et~al.}(2005){de Zotti}, {Ricci}, {Mesa}, {Silva},
  {Mazzotta}, {Toffolatti}, \& {Gonz{\'a}lez-Nuevo}}]{zotti2005}
{de Zotti}, G., {Ricci}, R., {Mesa}, D., {Silva}, L., {Mazzotta}, P.,
  {Toffolatti}, L., \& {Gonz{\'a}lez-Nuevo}, J. 2005, \aap, 431, 893

\bibitem[{{Dobbs} {et~al.}(2006){Dobbs}, {Halverson}, {Ade}, {Basu}, {Beelen},
  {Bertoldi}, {Cohalan}, {Cho}, {G{\"u}sten}, {Holzapfel}, {Kermish},
  {Kneissl}, {Kov{\'a}cs}, {Kreysa}, {Lanting}, {Lee}, {Lueker}, {Mehl},
  {Menten}, {Muders}, {Nord}, {Plagge}, {Richards}, {Schilke}, {Schwan},
  {Spieler}, {Weiss}, \& {White}}]{dobbs2006}
{Dobbs}, M., {Halverson}, N.~W., {Ade}, P.~A.~R., {Basu}, K., {Beelen}, A.,
  {Bertoldi}, F., {Cohalan}, C., {Cho}, H.~M., {G{\"u}sten}, R., {Holzapfel},
  W.~L., {Kermish}, Z., {Kneissl}, R., {Kov{\'a}cs}, A., {Kreysa}, E.,
  {Lanting}, T.~M., {Lee}, A.~T., {Lueker}, M., {Mehl}, J., {Menten}, K.~M.,
  {Muders}, D., {Nord}, M., {Plagge}, T., {Richards}, P.~L., {Schilke}, P.,
  {Schwan}, D., {Spieler}, H., {Weiss}, A., \& {White}, M. 2006, New Astronomy
  Review, 50, 960

\bibitem[{{Dunkley} {et~al.}(2009){Dunkley}, {Komatsu}, {Nolta}, {Spergel},
  {Larson}, {Hinshaw}, {Page}, {Bennett}, {Gold}, {Jarosik}, {Weiland},
  {Halpern}, {Hill}, {Kogut}, {Limon}, {Meyer}, {Tucker}, {Wollack}, \&
  {Wright}}]{dunkley2009}
{Dunkley}, J., {Komatsu}, E., {Nolta}, M.~R., {Spergel}, D.~N., {Larson}, D.,
  {Hinshaw}, G., {Page}, L., {Bennett}, C.~L., {Gold}, B., {Jarosik}, N.,
  {Weiland}, J.~L., {Halpern}, M., {Hill}, R.~S., {Kogut}, A., {Limon}, M.,
  {Meyer}, S.~S., {Tucker}, G.~S., {Wollack}, E., \& {Wright}, E.~L. 2009,
  \apjs, 180, 306

\bibitem[{{Dye} {et~al.}(2009){Dye}, {Ade}, {Bock}, {Chapin}, {Devlin},
  {Dunlop}, {Eales}, {Griffin}, {Gundersen}, {Halpern}, {Hargrave}, {Hughes},
  {Klein}, {Magnelli}, {Marsden}, {Mauskopf}, {Moncelsi}, {Netterfield},
  {Olmi}, {Pascale}, {Patanchon}, {Rex}, {Scott}, {Semisch}, {Thomas}, {Truch},
  {Tucker}, {Tucker}, {Viero}, \& {Wiebe}}]{dye2009}
{Dye}, S., {Ade}, P.~A.~R., {Bock}, J.~J., {Chapin}, E.~L., {Devlin}, M.~J.,
  {Dunlop}, J.~S., {Eales}, S.~A., {Griffin}, M., {Gundersen}, J.~O.,
  {Halpern}, M., {Hargrave}, P.~C., {Hughes}, D.~H., {Klein}, J., {Magnelli},
  B., {Marsden}, G., {Mauskopf}, P., {Moncelsi}, L., {Netterfield}, C.~B.,
  {Olmi}, L., {Pascale}, E., {Patanchon}, G., {Rex}, M., {Scott}, D.,
  {Semisch}, C., {Thomas}, N., {Truch}, M.~D.~P., {Tucker}, C., {Tucker},
  G.~S., {Viero}, M.~P., \& {Wiebe}, D.~V. 2009, ArXiv e-prints

\bibitem[{{Granato} {et~al.}(2004){Granato}, {De Zotti}, {Silva}, {Bressan}, \&
  {Danese}}]{granato2004}
{Granato}, G.~L., {De Zotti}, G., {Silva}, L., {Bressan}, A., \& {Danese}, L.
  2004, \apj, 600, 580

\bibitem[{{Greve} {et~al.}(2004){Greve}, {Ivison}, {Bertoldi}, {Stevens},
  {Dunlop}, {Lutz}, \& {Carilli}}]{greve2004}
{Greve}, T.~R., {Ivison}, R.~J., {Bertoldi}, F., {Stevens}, J.~A., {Dunlop},
  J.~S., {Lutz}, D., \& {Carilli}, C.~L. 2004, \mnras, 354, 779

\bibitem[{{G{\"u}sten} {et~al.}(2006){G{\"u}sten}, {Nyman}, {Schilke},
  {Menten}, {Cesarsky}, \& {Booth}}]{guesten2006}
{G{\"u}sten}, R., {Nyman}, L.~{\AA}., {Schilke}, P., {Menten}, K., {Cesarsky},
  C., \& {Booth}, R. 2006, \aap, 454, L13

\bibitem[{{Halverson} {et~al.}(2009){Halverson}, {Lanting}, {Ade}, {Basu},
  {Bender}, {Benson}, {Bertoldi}, {Cho}, {Chon}, {Clarke}, {Dobbs}, {Ferrusca},
  {Guesten}, {Holzapfel}, {Kovacs}, {Kennedy}, {Kermish}, {Kneissl}, {Lee},
  {Lueker}, {Mehl}, {Menten}, {Muders}, {Nord}, {Pacaud}, {Plagge},
  {Reichardt}, {Richards}, {Schaaf}, {Schilke}, {Schuller}, {Schwan},
  {Spieler}, {Tucker}, {Weiss}, \& {Zahn}}]{halverson08}
{Halverson}, N.~W., {Lanting}, T., {Ade}, P.~A.~R., {Basu}, K., {Bender},
  A.~N., {Benson}, B.~A., {Bertoldi}, F., {Cho}, H.~., {Chon}, G., {Clarke},
  J., {Dobbs}, M., {Ferrusca}, D., {Guesten}, R., {Holzapfel}, W.~L., {Kovacs},
  A., {Kennedy}, J., {Kermish}, Z., {Kneissl}, R., {Lee}, A.~T., {Lueker}, M.,
  {Mehl}, J., {Menten}, K.~M., {Muders}, D., {Nord}, M., {Pacaud}, F.,
  {Plagge}, T., {Reichardt}, C., {Richards}, P.~L., {Schaaf}, R., {Schilke},
  P., {Schuller}, F., {Schwan}, D., {Spieler}, H., {Tucker}, C., {Weiss}, A.,
  \& {Zahn}, O. 2009, \apj, 701, 42

\bibitem[{{Hill} {et~al.}(2009){Hill}, {Weiland}, {Odegard}, {Wollack},
  {Hinshaw}, {Larson}, {Bennett}, {Halpern}, {Page}, {Dunkley}, {Gold},
  {Jarosik}, {Kogut}, {Limon}, {Nolta}, {Spergel}, {Tucker}, \&
  {Wright}}]{hill08}
{Hill}, R.~S., {Weiland}, J.~L., {Odegard}, N., {Wollack}, E., {Hinshaw}, G.,
  {Larson}, D., {Bennett}, C.~L., {Halpern}, M., {Page}, L., {Dunkley}, J.,
  {Gold}, B., {Jarosik}, N., {Kogut}, A., {Limon}, M., {Nolta}, M.~R.,
  {Spergel}, D.~N., {Tucker}, G.~S., \& {Wright}, E.~L. 2009, \apjs, 180, 246

\bibitem[{{Hivon} {et~al.}(2002){Hivon}, {G{\'o}rski}, {Netterfield}, {Crill},
  {Prunet}, \& {Hansen}}]{hivon02}
{Hivon}, E., {G{\'o}rski}, K.~M., {Netterfield}, C.~B., {Crill}, B.~P.,
  {Prunet}, S., \& {Hansen}, F. 2002, \apj, 567, 2

\bibitem[{{Knox} {et~al.}(2004){Knox}, {Holder}, \& {Church}}]{knox2004}
{Knox}, L., {Holder}, G.~P., \& {Church}, S.~E. 2004, \apj, 612, 96

\bibitem[{{Komatsu} {et~al.}(2009){Komatsu}, {Dunkley}, {Nolta}, {Bennett},
  {Gold}, {Hinshaw}, {Jarosik}, {Larson}, {Limon}, {Page}, {Spergel},
  {Halpern}, {Hill}, {Kogut}, {Meyer}, {Tucker}, {Weiland}, {Wollack}, \&
  {Wright}}]{komatsu2008}
{Komatsu}, E., {Dunkley}, J., {Nolta}, M.~R., {Bennett}, C.~L., {Gold}, B.,
  {Hinshaw}, G., {Jarosik}, N., {Larson}, D., {Limon}, M., {Page}, L.,
  {Spergel}, D.~N., {Halpern}, M., {Hill}, R.~S., {Kogut}, A., {Meyer}, S.~S.,
  {Tucker}, G.~S., {Weiland}, J.~L., {Wollack}, E., \& {Wright}, E.~L. 2009,
  \apjs, 180, 330

\bibitem[{{Komatsu} \& {Seljak}(2002)}]{komatsu2002}
{Komatsu}, E. \& {Seljak}, U. 2002, \mnras, 336, 1256

\bibitem[{{Muhleman} \& {Berge}(1991)}]{muhleman91}
{Muhleman}, D.~O. \& {Berge}, G.~L. 1991, Icarus, 92, 263

\bibitem[{{Negrello} {et~al.}(2007){Negrello}, {Perrotta},
  {Gonz{\'a}lez-Nuevo}, {Silva}, {de Zotti}, {Granato}, {Baccigalupi}, \&
  {Danese}}]{negrello2007}
{Negrello}, M., {Perrotta}, F., {Gonz{\'a}lez-Nuevo}, J., {Silva}, L., {de
  Zotti}, G., {Granato}, G.~L., {Baccigalupi}, C., \& {Danese}, L. 2007,
  \mnras, 377, 1557

\bibitem[{{Pacaud} {et~al.}(2007){Pacaud}, {Pierre}, {Adami}, {Altieri},
  {Andreon}, {Chiappetti}, {Detal}, {Duc}, {Galaz}, {Gueguen}, {Le F{\`e}vre},
  {Hertling}, {Libbrecht}, {Melin}, {Ponman}, {Quintana}, {Refregier},
  {Sprimont}, {Surdej}, {Valtchanov}, {Willis}, {Alloin}, {Birkinshaw},
  {Bremer}, {Garcet}, {Jean}, {Jones}, {Le F{\`e}vre}, {Maccagni}, {Mazure},
  {Proust}, {R{\"o}ttgering}, \& {Trinchieri}}]{pacaud2007}
{Pacaud}, F., {Pierre}, M., {Adami}, C., {Altieri}, B., {Andreon}, S.,
  {Chiappetti}, L., {Detal}, A., {Duc}, P.-A., {Galaz}, G., {Gueguen}, A., {Le
  F{\`e}vre}, J.-P., {Hertling}, G., {Libbrecht}, C., {Melin}, J.-B., {Ponman},
  T.~J., {Quintana}, H., {Refregier}, A., {Sprimont}, P.-G., {Surdej}, J.,
  {Valtchanov}, I., {Willis}, J.~P., {Alloin}, D., {Birkinshaw}, M., {Bremer},
  M.~N., {Garcet}, O., {Jean}, C., {Jones}, L.~R., {Le F{\`e}vre}, O.,
  {Maccagni}, D., {Mazure}, A., {Proust}, D., {R{\"o}ttgering}, H.~J.~A., \&
  {Trinchieri}, G. 2007, \mnras, 382, 1289

\bibitem[{{Pierre} {et~al.}(2004){Pierre}, {Valtchanov}, {Altieri}, \& {et
  al.}}]{pierre2004}
{Pierre}, M., {Valtchanov}, I., {Altieri}, B., \& {et al.} 2004, Journal of
  Cosmology and Astro-Particle Physics, 9, 11

\bibitem[{{QUaD collaboration: R.~B.~Friedman} {et~al.}(2009){QUaD
  collaboration: R.~B.~Friedman}, {Ade}, {Bock}, {Bowden}, {Brown}, {Cahill},
  {Castro}, {Church}, {Culverhouse}, {Ganga}, {Gear}, {Gupta}, {Hinderks},
  {Kovac}, {Lange}, {Leitch}, {Melhuish}, {Memari}, {Murphy}, {Orlando}, {O'
  Sullivan}, {Piccirillo}, {Pryke}, {Rajguru}, {Rusholme}, {Schwarz}, {Taylor},
  {Thompson}, {Turner}, {Wu}, \& {Zemcov}}]{friedman2009}
{QUaD collaboration: R.~B.~Friedman}, {Ade}, P., {Bock}, J., {Bowden}, M.,
  {Brown}, M.~L., {Cahill}, G., {Castro}, P.~G., {Church}, S., {Culverhouse},
  T., {Ganga}, K., {Gear}, W.~K., {Gupta}, S., {Hinderks}, J., {Kovac}, J.,
  {Lange}, A.~E., {Leitch}, E., {Melhuish}, S.~J., {Memari}, Y., {Murphy},
  J.~A., {Orlando}, A., {O' Sullivan}, C., {Piccirillo}, L., {Pryke}, C.,
  {Rajguru}, N., {Rusholme}, B., {Schwarz}, R., {Taylor}, A.~N., {Thompson},
  K.~L., {Turner}, A.~H., {Wu}, E.~Y.~S., \& {Zemcov}, M. 2009, \apj, 700, L187

\bibitem[{{Reichardt} {et~al.}(2009){Reichardt}, {Ade}, {Bock}, {Bond},
  {Brevik}, {Contaldi}, {Daub}, {Dempsey}, {Goldstein}, {Holzapfel}, {Kuo},
  {Lange}, {Lueker}, {Newcomb}, {Peterson}, {Ruhl}, {Runyan}, \&
  {Staniszewski}}]{reichardt2008}
{Reichardt}, C.~L., {Ade}, P.~A.~R., {Bock}, J.~J., {Bond}, J.~R., {Brevik},
  J.~A., {Contaldi}, C.~R., {Daub}, M.~D., {Dempsey}, J.~T., {Goldstein},
  J.~H., {Holzapfel}, W.~L., {Kuo}, C.~L., {Lange}, A.~E., {Lueker}, M.,
  {Newcomb}, M., {Peterson}, J.~B., {Ruhl}, J., {Runyan}, M.~C., \&
  {Staniszewski}, Z. 2009, \apj, 694, 1200

\bibitem[{{Rudy} {et~al.}(1987){Rudy}, {Muhleman}, {Berge}, {Jakosky}, \&
  {Christensen}}]{rudy87}
{Rudy}, D.~J., {Muhleman}, D.~O., {Berge}, G.~L., {Jakosky}, B.~M., \&
  {Christensen}, P.~R. 1987, Icarus, 71, 159

\bibitem[{{Sayers} {et~al.}(2009){Sayers}, {Golwala}, {Rossinot}, {Ade},
  {Aguirre}, {Bock}, {Edgington}, {Glenn}, {Goldin}, {Haig}, {Lange},
  {Laurent}, {Mauskopf}, \& {Nguyen}}]{sayers09}
{Sayers}, J., {Golwala}, S.~R., {Rossinot}, P., {Ade}, P.~A.~R., {Aguirre},
  J.~E., {Bock}, J.~J., {Edgington}, S.~F., {Glenn}, J., {Goldin}, A., {Haig},
  D., {Lange}, A.~E., {Laurent}, G.~T., {Mauskopf}, P.~D., \& {Nguyen}, H.~T.
  2009, \apj, 690, 1597

\bibitem[{{Schwan} {et~al.}(2003){Schwan}, {Bertoldi}, {Cho}, {Dobbs},
  {Guesten}, {Halverson}, {Holzapfel}, {Kreysa}, {Lanting}, {Lee}, {Lueker},
  {Mehl}, {Menten}, {Muders}, {Myers}, {Plagge}, {Raccanelli}, {Schilke},
  {Richards}, \& {White}}]{schwan2003}
{Schwan}, D., {Bertoldi}, F., {Cho}, S., {Dobbs}, M., {Guesten}, R.,
  {Halverson}, N.~W., {Holzapfel}, W.~L., {Kreysa}, E., {Lanting}, T.~M.,
  {Lee}, A.~T., {Lueker}, M., {Mehl}, J., {Menten}, K., {Muders}, D., {Myers},
  M., {Plagge}, T., {Raccanelli}, A., {Schilke}, P., {Richards}, P.~L. an
  d~{Spieler}, H., \& {White}, M. 2003, New Astronomy Review, 47, 933

\bibitem[{Schwan {et~al.}(2009)}]{schwan2008}
Schwan, D. {et~al.} 2009, in preparation

\bibitem[{{Sharp} {et~al.}(2009){Sharp}, {Marrone}, {Carlstrom}, {Culverhouse},
  {Greer}, {Hawkins}, {Hennessy}, {Joy}, {Lamb}, {Leitch}, {Loh}, {Miller},
  {Mroczkowski}, {Muchovej}, {Pryke}, \& {Woody}}]{sharp2009}
{Sharp}, M.~K., {Marrone}, D.~P., {Carlstrom}, J.~E., {Culverhouse}, T.,
  {Greer}, C., {Hawkins}, D., {Hennessy}, R., {Joy}, M., {Lamb}, J.~W.,
  {Leitch}, E.~M., {Loh}, M., {Miller}, A., {Mroczkowski}, T., {Muchovej}, S.,
  {Pryke}, C., \& {Woody}, D. 2009, ArXiv e-prints; astro-ph/0901.4342

\bibitem[{{Shaw} {et~al.}(2009){Shaw}, {Zahn}, {Holder}, \&
  {Dor{\'e}}}]{shaw2009}
{Shaw}, L.~D., {Zahn}, O., {Holder}, G.~P., \& {Dor{\'e}}, O. 2009, ArXiv
  e-prints, astro-ph/0903.5322

\bibitem[{{Sievers} {et~al.}(2009){Sievers}, {Mason}, {Weintraub}, {Achermann},
  {Altamirano}, {Bond}, {Bronfman}, {Bustos}, {Contaldi}, {Dickinson}, {Jones},
  {May}, {Myers}, {Oyarce}, {Padin}, {Pearson}, {Pospieszalski}, {Readhead},
  {Reeves}, {Shepherd}, {Taylor}, \& {Torres}}]{sievers2009}
{Sievers}, J.~L., {Mason}, B.~S., {Weintraub}, L., {Achermann}, C.,
  {Altamirano}, P., {Bond}, J.~R., {Bronfman}, L., {Bustos}, R., {Contaldi},
  C., {Dickinson}, C., {Jones}, M.~E., {May}, J., {Myers}, S.~T., {Oyarce}, N.,
  {Padin}, S., {Pearson}, T.~J., {Pospieszalski}, M., {Readhead}, A.~C.~S.,
  {Reeves}, R., {Shepherd}, M.~C., {Taylor}, A.~C., \& {Torres}, S. 2009, ArXiv
  e-prints, astro-ph/0901.4540

\bibitem[{{Sunyaev} \& {Zel'dovich}(1972)}]{sunyaev72}
{Sunyaev}, R.~A. \& {Zel'dovich}, Y.~B. 1972, Comments on Astrophysics, 4, 173

\bibitem[{{Viero} {et~al.}(2009){Viero}, {Ade}, {Bock}, {Chapin}, {Devlin},
  {Griffin}, {Gundersen}, {Halpern}, {Hargrave}, {Hughes}, {Klein},
  {MacTavish}, {Marsden}, {Martin}, {Mauskopf}, {Moncelsi}, {Negrello},
  {Netterfield}, {Olmi}, {Pascale}, {Patanchon}, {Rex}, {Scott}, {Semisch},
  {Thomas}, {Truch}, {Tucker}, {Tucker}, \& {Wiebe}}]{viero2009}
{Viero}, M.~P., {Ade}, P.~A.~R., {Bock}, J.~J., {Chapin}, E.~L., {Devlin},
  M.~J., {Griffin}, M., {Gundersen}, J.~O., {Halpern}, M., {Hargrave}, P.~C.,
  {Hughes}, D.~H., {Klein}, J., {MacTavish}, C.~J., {Marsden}, G., {Martin},
  P.~G., {Mauskopf}, P., {Moncelsi}, L., {Negrello}, M., {Netterfield}, C.~B.,
  {Olmi}, L., {Pascale}, E., {Patanchon}, G., {Rex}, M., {Scott}, D.,
  {Semisch}, C., {Thomas}, N., {Truch}, M.~D.~P., {Tucker}, C., {Tucker},
  G.~S., \& {Wiebe}, D.~V. 2009, ArXiv e-prints, astro-ph/0904.1200

\bibitem[{{Vikhlinin} {et~al.}(2009){Vikhlinin}, {Kravtsov}, {Burenin},
  {Ebeling}, {Forman}, {Hornstrup}, {Jones}, {Murray}, {Nagai}, {Quintana}, \&
  {Voevodkin}}]{vikhlinin2009}
{Vikhlinin}, A., {Kravtsov}, A.~V., {Burenin}, R.~A., {Ebeling}, H., {Forman},
  W.~R., {Hornstrup}, A., {Jones}, C., {Murray}, S.~S., {Nagai}, D.,
  {Quintana}, H., \& {Voevodkin}, A. 2009, \apj, 692, 1060

\bibitem[{{Willis} {et~al.}(2005){Willis}, {Pacaud}, {Valtchanov}, {Pierre},
  {Ponman}, {Read}, {Andreon}, {Altieri}, {Quintana}, {Dos Santos},
  {Birkinshaw}, {Bremer}, {Duc}, {Galaz}, {Gosset}, {Jones}, \&
  {Surdej}}]{willis2005}
{Willis}, J.~P., {Pacaud}, F., {Valtchanov}, I., {Pierre}, M., {Ponman}, T.,
  {Read}, A., {Andreon}, S., {Altieri}, B., {Quintana}, H., {Dos Santos}, S.,
  {Birkinshaw}, M., {Bremer}, M., {Duc}, P.-A., {Galaz}, G., {Gosset}, E.,
  {Jones}, L., \& {Surdej}, J. 2005, \mnras, 363, 675

\bibitem[{{Zhang} {et~al.}(2006){Zhang}, {B{\"o}hringer}, {Finoguenov},
  {Ikebe}, {Matsushita}, {Schuecker}, {Guzzo}, \& {Collins}}]{zhang2006}
{Zhang}, Y.-Y., {B{\"o}hringer}, H., {Finoguenov}, A., {Ikebe}, Y.,
  {Matsushita}, K., {Schuecker}, P., {Guzzo}, L., \& {Collins}, C.~A. 2006,
  \aap, 456, 55

\end{thebibliography}

\end{document}